\documentclass{fundam} % Fundamenta Informatica Document Style
\publyear{25}
\papernumber{3}
\volume{195}
\issue{1-4}
\theDOI{10.46298/fi.11710}

\versionForARXIV

\usepackage{bm} % Allows boldface math
\usepackage{hyperref} % Takes care of hyperlinks
\usepackage{graphicx} % Allows FO3COR.drawio.pdf to be used
\usepackage{listings} % Used to display Python code nicely

\def\full{\bm{T}} % Define symbol for the universal relation
\def\empty{\bm{0}} % Define symbol for the empty relation
\def\identity{\bm{1}} % Define symbol for the identity relation
\def\true{\bm{t}} % Define symbol for the boolean value True
\def\false{\bm{f}} % Define symbol for the boolean value False

\begin{document}

\title{Translating Three-Variable First-Order Predicate Logic to Relation Algebra, Implemented using Z3}

\author{Anthony Brogni \thanks{Address for correspondence: brogn002@umn.edu} \\
Department of Computer Science and Engineering \\
University of Minnesota Twin Cities \\
Minneapolis, Minnesota, United States \\
brogn002@umn.edu \\
\and Sebastiaan J. C. Joosten \\
Department of Computer Science and Engineering\thanks{Former institution at which this research was carried out.} \\
University of Minnesota Twin Cities \\
Minneapolis, Minnesota, United States \\
sjoosten@amazon.com}

\maketitle

\runninghead{A. Brogni, S.J.C. Joosten}{Translating Three-Variable First-Order Predicate Logic to Relation Algebra}

\vspace*{-10ex}

\begin{abstract}
This paper presents the development of a software tool that enables the translation of first-order predicate logic with at most three variables into relation algebra. The tool was developed using the Z3 theorem prover, leveraging its capabilities to enhance reliability, generate code, and expedite development. The resulting standalone Python program allows users to translate first-order logic formulas into relation algebra, eliminating the need to work with relation algebra explicitly. This paper outlines the theoretical background of first-order logic, relation algebra, and the translation process. It also describes the implementation details, including validation of the software tool using Z3 for testing correctness. By demonstrating the feasibility of utilizing first-order logic as an alternative language for expressing relation algebra, this tool paves the way for integrating first-order logic into tools traditionally relying on relation algebra as input.
\end{abstract}


\section{Introduction}
In the past, multiple tools have been developed that provide the convenience of relation algebra for its succinct expressive power as an input language. However, what the RAMiCS community knows as an advantage, can simultaneously hinder the uptake of these tools. In our development of Ampersand~\cite{Joosten2015}, we found that some people choose not to use the tool because it would require them to learn how to express their ideas in Relation Algebra. In this paper, we provide a gateway tool that translates first-order logic into relation algebra (and back). The tool developed might interest those developing or using tools with relation algebra as their input language. This paper additionally describes the development process, which includes methods for early validation during its development. This process might interest tool developers more widely, as the methods might apply to any tool that handles an algebra that can translate into first-order logic.

In the process of developing this tool, we used the theorem prover Z3~\cite{z3_theorem_prover}, developed by Microsoft Research, to increase reliability, generate part of the code, and speed up development. The final software product is a stand-alone Python tool that does not depend on Z3. This Python implementation addresses an argument we occasionally encounter in our development of Ampersand~\cite{Joosten2015}: some people choose not to adopt the tool because it would require them to learn how to express their ideas in Relation Algebra. By providing an automatic translation from first-order logic, they do not have to. We believe that some of the ideas used to write this code can be applied in the development of other formal tools as well.

Relation Algebraic operators are often defined in terms of First-Order Logic, so the translation of Relation Algebra into equivalent First-Order Logic formulas is well known. It is also known that one can go in the other direction for First-Order Logic formulas that concern only binary relations and use no more than three variables. We follow the procedure by Yoshiki Nakamura~\cite{Nakamura2020} in our implementation. This procedure outlines how to translate a formula in First-Order Logic with no more than three variables (an FO3 formula) into one in Relation Algebra (RA). However, the domain of all predicates and quantifiers is the same in this translation. Tools that use Relation Algebra as an input language like Ampersand and RelView~\cite{Berghammer2005} use heterogeneous Relation Algebra.

Consequently, we adapt the translation procedure to translate FO3 formulas with quantifiers over different domains into well-typed heterogeneous Relation Algebra. The translation procedure is implemented as a stand-alone Python program that can be run independently so it can be useful for any tool that takes Relation Algebra as its input language. The implementation is provided as an artifact to this paper~\cite{doi_of_implementation}. We took care to make the implementation highly readable and to encourage re-implementation into other tools as well. This paper describes the adaptations we needed to make to the procedure to support heterogeneous relation algebra, as well as how we ensured the correctness of our implementation.

\begin{figure}[htbp]
  \centering
  \includegraphics{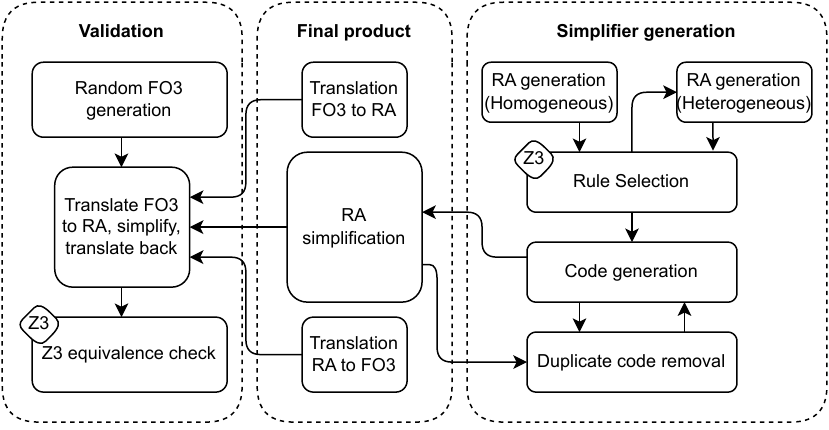}
  \caption{The Development Process. Arrows signify the flow of information. Boxes are software components.}
  \label{fig:development}
\end{figure}

Figure~\ref{fig:development} summarizes our development process. The final product consists of three key components: a function to convert FO3 formulas into RA, one to convert back, and a simplification tool for reducing RA formulas. The translation tools are developed manually and Z3 validates its correctness. In contrast, Z3 assumes a more direct role in creating the simplification tool. We wrote code to generate all potential simplification rules of increasing size, and rely on Z3 to check their validity, then store the valid rules in a dictionary. Another Python program uses this rule dictionary to generate Python code for simplifying RA formulas. The resulting simplifier is also checked (redundantly) by Z3, but its source code is quite large and it would have been tedious to get it right by hand.

All components of this tool come in two variants: one for homogeneous relation algebra and first-order logic with a single sort, and one for heterogeneous relation algebra and first-order logic with multiple sorts.

\subsection{Key Contributions}
The key contributions of this paper include:
\begin{itemize}
\item An implementation of a translation from FO3 to RA, both for untyped and typed FO3, translating into homogeneous and heterogeneous RA, and back.
\item A method of using Z3 during the development of formal tools and in a testing setup.
\item A simplifier for RA, which plays a key role in our translation by producing succinct formulas.
\item A method for combining Z3 and code generation to write a correct simplifier.
\end{itemize}

% Create Figure 1

\section{Theoretical Background}\label{background}
In the following, we assume a global set of non-empty sorts $\mathcal{D}$, a global set of binary predicate symbols $\mathcal{A}$, and a function $d : \mathcal{A} \rightarrow \mathcal{D} \times \mathcal{D}$ that denotes the type of each predicate symbol, where $d_1$ and $d_2$ are used to denote the two components of $d$: $d(a) = (d_1(a), d_2(a))$.

An FO3 formula is defined over the following language, where $a \in \mathcal{A}$ is from the set of binary predicate symbols, $x, y$ are from a set of three variables, and $D \in \mathcal{D}$ is from the set of sorts:
\[\varphi, \psi \in \text{FO3} = a(x, y) \mid x = y \mid \true \mid \false \mid \varphi \vee \psi \mid \varphi \wedge \psi \mid \exists x \in D.~ \varphi \mid \forall x \in D.~ \varphi \mid \neg \varphi\]

We say an FO3 formula is closed if every variable occurring in it is bound by a quantifier ($\forall$ or~$\exists$). We say that the type of an occurrence of a variable in a closed formula is the domain $D$ specified in the quantifier that binds it. We say that an FO3 formula is well-typed if, for every occurrence of a predicate $a(x, y)$, the types of $x$ and $y$ match $d_1(a)$ and $d_2(a)$, respectively.

If $\mathcal{D}$ has precisely one element, we call it the universal set $\mathcal{U}$ and say that the language is homogeneous (corresponding to single-sorted first-order logic in the case of FO3). If we place no such restriction on $\mathcal{D}$, we call the language heterogeneous (corresponding to many-sorted first-order logic).

This detailed structure of the language definition is maintained to ensure clarity and consistency with related works, such as Nakamura's. Retaining this structure avoids potential ambiguities.

\vspace{1em} % This produces a blank line

An RA formula is a formula over the following language, where $a$ is from $\mathcal{A}$, and $D_1, D_2$ from $\mathcal{D}$:
\[\varphi, \psi \in \text{RA} = a[D_1, D_2] \mid \full[D_1, D_2] \mid \empty[D_1, D_2] \mid \identity[D_1] \mid \varphi \cup \psi \mid \varphi \cap \psi \mid \varphi \circ \psi \mid \varphi \dagger \psi \mid \overline{\varphi} \mid \varphi^{-1}\]

We overload $d$ to denote the type of an RA formula by defining it as follows:
\[d(a[D_1, D_2]) = (D_1, D_2), \quad d(\full[D_1, D_2]) = (D_1, D_2),\] \[d(\empty[D_1, D_2]) = (D_1, D_2), \quad d(\identity[D_1]) = (D_1, D_1),\]
\[d(\varphi \cup \psi) = d(\varphi), \quad d(\varphi \cap \psi) = d(\varphi), \quad d(\varphi \circ \psi) = (d_1(\varphi), d_2(\psi)), \quad d(\varphi \dagger \psi) = (d_1(\varphi), d_2(\psi)),\]
\[d(\overline{\varphi}) = d(\varphi), \quad d(\varphi^{-1}) = (d_2(\varphi), d_1(\varphi))\]

Each binary predicate symbol $a$ has a fixed type denoted by $d(a) = (D_1, D_2)$. That is, if $a[D_1, D_2]$ is used in a formula, then $d(a) = (D_1, D_2)$ must hold for all occurrences of the binary predicate symbol $a$ in the formula, so $a[D_2, D_1]$ for instance would not be valid. Thus, we say that a formula in RA is well-typed if all of the following conditions hold:
\begin{enumerate}
    \item Every occurrence of the same predicate symbol $a$ has the same fixed type $d(a)$.
    \item Occurrences of $\varphi \cup \psi$ and $\varphi \cap \psi$ satisfy $d(\varphi) = d(\psi)$.
    \item Occurrences of $\varphi \circ \psi$ and $\varphi \dagger \psi$ satisfy $d_2(\varphi) = d_1(\psi)$.
\end{enumerate}

To describe the semantics of FO3 and RA, we use interpretation functions $\mathcal{I}_\mathcal{A} : \mathcal{A} \times \mathcal{D} \times \mathcal{D} \to \mathcal{P}(S\times S)$ and $\mathcal{I}_\mathcal{D} : \mathcal{D} \to (\mathcal{P}(S)-\{\emptyset\})$ for some set of constants $S$ (disjoint from the set of three variables) subject to the typing constraint: $\mathcal{I}_\mathcal{A}(a[D_1, D_2]) \subseteq \mathcal{I}_\mathcal{D}(D_1)\times\mathcal{I}_\mathcal{D}(D_2)$. Note that each $a[D_1, D_2]$ is a tuple $(a, D_1, D_2)$ in $\mathcal{A} \times \mathcal{D} \times \mathcal{D}$. To take care of variables in FO3 formulas, define FO3+ as FO3 with the additional relaxation that we can substitute variables for constants in $S$. We then define $\mathcal{I}$ as a function that takes a closed formula in FO3+, a formula in RA, or a domain and produces a Boolean, a set of pairs, or a set, respectively. On a domain, $\mathcal{I}(D) = \mathcal{I}_\mathcal{D}(D)$. On closed formulas in FO3+, define $\mathcal{I}(a(x',y'))$ as true if and only if $(x',y') \in \mathcal{I}_\mathcal{A}(a[D_1, D_2])$ (where $x'$ and $y'$ are constants in $S$, possibly the same elements) and let the other operations have their usual meaning (i.e. $\mathcal{I}(\phi \vee \psi) = \mathcal{I}(\phi) \vee \mathcal{I}(\psi)$ and so on).

For formulas in RA, let:
\begin{align*}
 \mathcal{I}(a[D_1, D_2]) &= \mathcal{I}_\mathcal{A}(a[D_1, D_2]) \\
 \mathcal{I}(\full[D_1, D_2]) &= \mathcal{I}_\mathcal{D}(D_1) \times \mathcal{I}_\mathcal{D}(D_2) \\
 \mathcal{I}(\empty[D_1, D_2]) &= \emptyset \\
 \mathcal{I}(\identity[D_1]) &= \{(x, x) \mid x \in \mathcal{I}_\mathcal{D}(D_1)\} \\
 \mathcal{I}(\varphi \cup \psi) &= \{(x, y) \mid (x, y) \in \mathcal{I}(\varphi) \lor (x, y) \in \mathcal{I}(\psi)\} \\
 \mathcal{I}(\varphi \cap \psi) &= \{(x, y) \mid (x, y) \in \mathcal{I}(\varphi) \land (x, y) \in \mathcal{I}(\psi)\}  \\
 \mathcal{I}(\varphi \circ \psi) &= \{(x, y) \mid \exists z.~ (x, z)\in \mathcal{I}(\varphi) \land (z, y) \in \mathcal{I}(\psi)\} \\
 \mathcal{I}(\varphi \dagger \psi) &= \{(x, y) \mid \forall z \in \mathcal{I}_\mathcal{D}(d_2(\varphi)).~ (x, z)\in \mathcal{I}(\varphi) \lor (z, y) \in \mathcal{I}(\psi)\} \\
 \mathcal{I}(\overline{\varphi}) &= \{(x, y) \mid x \in \mathcal{I}_\mathcal{D}(d_1(\varphi)) \wedge y \in \mathcal{I}_\mathcal{D}(d_2(\varphi))\ \wedge\  (x, y) \notin \mathcal{I}(\varphi)\} \\
 \mathcal{I}(\varphi^{-1})&= \{(x, y) \mid (y, x) \in \mathcal{I}(\varphi)\}
\end{align*}
If $\varphi$ is a well-typed closed formula in FO3 and $\psi$ is a well-typed formula in RA, we say that the two are equivalent if for all choices of $\mathcal{I}_\mathcal{A}, \mathcal{I}_\mathcal{D}$ and $S$, the value $\mathcal{I}(\varphi)$ is true if and only if $\mathcal{I}(\psi) = \{(x, y) \mid x\in d_1(\psi) ~\land~ y\in d_2(\psi)\}$. We then say that a translation of $\varphi$ in FO3 into RA is sound if it and its translation are equivalent.

\subsection{The Homogeneous Translation Process}\label{translation}
In the translation process outlined by Nakamura, there are four steps~\cite{Nakamura2020}. In this section, we follow that translation for the special case in which the logic is homogeneous. To achieve full correspondence, we abbreviate $a = a[\mathcal{U},\mathcal{U}]$, $\full = \full[\mathcal{U},\mathcal{U}]$, $\empty = \empty[\mathcal{U},\mathcal{U}]$, and $\identity = \identity[\mathcal{U}]$.

The first three steps add properties to the formula that are preserved by subsequent steps. First, applying De Morgan's laws puts formulas into negation normal form.

Second, the $\wedge$ and $\vee$ operators are distributed such that each $\exists$ contains a formula in conjunctive normal form, and each $\forall$ contains a formula in disjunctive normal form. Nakamura refers to this as a `good' FO3 formula. More formally, the translation from FO3 formulas in negation normal form into `good' FO3 formulas is defined below, where the translation of $\varphi$ is obtained from $\text{T}_{-}(\varphi)$:
\begin{align*}
    \text{T}_{\bullet}(\varphi) :=
        \begin{cases} 
            \varphi \quad (\bullet = -) \\
            \{\{\varphi\}\} \quad (\bullet = \exists, \forall)
        \end{cases} \text{if $\varphi$ is an atomic or negated atomic formula.} \\
    \text{T}_{\bullet}(\exists z.\varphi) :=
        \begin{cases} 
            \bigvee_{i \in [n]} \exists z. \bigwedge \phi_i \quad (\bullet = -) \\
            \{\{\bigvee_{i \in [n]} \exists z. \bigwedge \phi_i\}\} \quad (\bullet = \exists, \forall)
        \end{cases} \text{where $\text{T}_{\exists}(\varphi) = \{ \phi_{i} \mid i \in [n]\}$.} \\
    \text{T}_{\bullet}(\forall z.\varphi) :=
        \begin{cases} 
            \bigwedge_{i \in [n]} \forall z. \bigvee \phi_i \quad (\bullet = -) \\
            \{\{\bigwedge_{i \in [n]} \forall z. \bigvee \phi_i\}\} \quad (\bullet = \exists, \forall)
        \end{cases} \text{where $\text{T}_{\forall}(\varphi) = \{ \phi_{i} \mid i \in [n]\}$.} 
\end{align*}
\begin{align*}
    \text{T}_{\bullet}(\psi_1 \land \psi_2) :=
        \begin{cases} 
            \text{T}_{-}(\psi_1) \land \text{T}_{-}(\psi_2) \quad (\bullet = -) \\
            \text{T}_{\forall}(\psi_1) \cup \text{T}_{\forall}(\psi_2) \quad (\bullet = \forall) \\
            \{ \Psi_1 \cup \Psi_2 \mid \Psi_1 \in \text{T}_{\exists}(\psi_1), \Psi_2 \in \text{T}_{\exists}(\psi_2) \} \quad (\bullet = \exists)
        \end{cases} \\
    \text{T}_{\bullet}(\psi_1 \lor \psi_2) :=
        \begin{cases} 
            \text{T}_{-}(\psi_1) \lor \text{T}_{-}(\psi_2) \quad (\bullet = -) \\
            \text{T}_{\exists}(\psi_1) \cup \text{T}_{\exists}(\psi_2) \quad (\bullet = \exists) \\
            \{ \Psi_1 \cup \Psi_2 \mid \Psi_1 \in \text{T}_{\forall}(\psi_1), \Psi_2 \in \text{T}_{\forall}(\psi_2) \} \quad (\bullet = \forall)
        \end{cases}
\end{align*}
For brevity, we use $\exists z.\varphi$ to express $\exists z \in D.~ \varphi$ and $\forall z.\varphi$ to express $\forall z \in D.~ \varphi$. If desired, a proof of the above translation and additional explanation are given in Nakamura's paper~\cite{Nakamura2020}.

Third, quantifiers are pushed inwards as much as possible by taking out formulas that do not depend on the variable being quantified over. It is important to note that each set $\mathcal{I}(D)$ is assumed to be non-empty; if $\mathcal{I}(D)$ were empty, we could not push quantifiers straightforwardly. The result is a formula where every existential quantifier, for instance over a variable $z$, contains a conjunction of formulas such that the conjuncts dependent on one variable (e.g., $x$) do not also depend on another variable (e.g., $y$). Similarly, every universal quantifier contains a disjunction of formulas with that property. This is referred to as a `nice' FO3 formula by Nakamura. More formally, the translation from `good' FO3 formulas into `nice' FO3 formulas is defined below, where the nice translation of $\varphi$ is $\text{T}(\varphi)$. A proof of correctness of this translation and additional explanation are given in Nakamura's paper~\cite{Nakamura2020}. For completeness, we list the translation below. Let $\psi_1$, $\psi_2$, and $\psi_3$ be formulas that depend on a subset of the variables $\{ x,y \}$, $\{ x,z \}$, and $\{ y,z \}$, respectively:
\begin{align*}
    \text{T}(\varphi) &:= \varphi \quad \text{if $\varphi$ is an atomic or negated atomic formula.} \\
    \text{T}(\psi \lor \rho) &:= \text{T}(\psi) \lor \text{T}(\rho) \\
    \text{T}(\psi \land \rho) &:= \text{T}(\psi) \land \text{T}(\rho) \\
    \text{T}(\exists z. \psi_1 \land \psi_2 \land \psi_3) &:= \text{T}(\psi_1) \land \exists z.\text{T}(\psi_2) \land \text{T}(\psi_3) \\
    \text{T}(\forall z. \psi_1 \lor \psi_2 \lor \psi_3) &:= \text{T}(\psi_1) \lor \forall z.\text{T}(\psi_2) \lor \text{T}(\psi_3)
\end{align*}

The final step of the translation procedure is to translate a `nice' FO3 formula into RA. The translation function we implemented for this final step, T, is given as:
\begin{align*}
    &\begin{aligned}
        \text{T}([\true]_{x,y}) &:= \full &
        \text{T}([\false]_{x,y}) &:= \empty \\
        \text{T}([a(x, y)]_{x,y}) &:= a &
        \text{T}([a(y, x)]_{x,y}) &:= a^{-1} \\
        \text{T}([a(x, x)]_{x,y}) &:= (a \cap \identity) \circ \full &
        \text{T}([a(y, y)]_{x,y}) &:= \full \circ (a \cap \identity) \\
        \text{T}([\neg(\varphi^{\{x, y\}})]_{x,y}) &:= \overline{\text{T}([\varphi^{\{x, y\}}]_{x,y})} &
        \text{T}([x = x]_{x,y}) &:= \full \\
        \text{T}([x = y]_{x,y}) &:= \identity &
        \text{T}([x = y]_{y,x}) &:= \identity^{-1} \end{aligned} \\ & \begin{aligned}
        \text{T}([\varphi^{\{x, y\}} \land \psi^{\{x, y\}}]_{x,y}) &:= \text{T}([\varphi^{\{x, y\}}]_{x,y}) \cap \text{T}([\psi^{\{x, y\}}]_{x,y}) \\
        \text{T}([\varphi^{\{x, y\}} \lor \psi^{\{x, y\}}]_{x,y}) &:= \text{T}([\varphi^{\{x, y\}}]_{x,y}) \cup \text{T}([\psi^{\{x, y\}}]_{x,y}) \\
        \text{T}([\exists z. \varphi^{\{x, z\}} \land \psi^{\{z, y\}}]_{x,y}) &:= \text{T}([\varphi^{\{x, z\}}]_{x,z}) \circ \text{T}([\psi^{\{z, y\}}]_{z,y}) \\
        \text{T}([\forall z. \varphi^{\{x, z\}} \lor \psi^{\{z, y\}}]_{x,y}) &:= \text{T}([\varphi^{\{x, z\}}]_{x,z}) \dagger \text{T}([\psi^{\{z, y\}}]_{z,y})
    \end{aligned}
\end{align*}
The notation being used for this translation function is $\text{T}([\varphi]_{a,b})$ where $\varphi$ is the FO3 formula to translate, and $a$ and $b$ are two arbitrary variables that are used to keep track of the source and target of the relation algebra formula being built by the translation. In our code~\cite{doi_of_implementation}, the translation function is thus implemented to accept three arguments: $\varphi$, $a$, and $b$. It is important to note that $a$ and $b$ must be pairwise distinct variables, otherwise the definition becomes ill-defined in the case of atomic formulas and equality.

In the last two lines of the translation function definition, $\varphi^{\{x, z\}}$ and $\psi^{\{z, y\}}$ stand for the conjuncts (disjuncts) that depend on the variables $\{x,z\}$ and $\{z,y\}$ inside the existential (universal) quantifier, respectively.

To demonstrate the homogeneous translation procedure, a fully worked example of the translation of the closed FO3 formula $\neg(\forall x. \forall y. (\neg(A(x,x))) \lor (\neg(A(y,y))))$ into RA is shown below:

\begin{enumerate}
    \item Original Expression: $\neg(\forall x. \forall y. (\neg(A(x,x))) \lor (\neg(A(y,y))))$
    \item Negation Normal Form: $\exists x. \exists y. (A(x,x)) \land (A(y,y))$
    \item Good FO3 Translation: $\exists x. \exists y. (A(x,x)) \land (A(y,y))$
    \item Nice FO3 Translation: $(\exists y. A(y,y)) \land (\exists x. A(x,x))$
    \item Final Translation: $((\full) \circ (((A) \cap (\identity)) \circ (\full))) \cap ((\full) \circ (((A) \cap (\identity)) \circ (\full)))$
    \item Final Translation Simplified: $(\full) \circ (((A) \cap (\identity)) \circ (\full))$
\end{enumerate}

\subsection{The Heterogeneous Translation Process}

The first three steps of the heterogeneous translation process are the same as those of the homogeneous translation process, described above in Section~\ref{translation}. However, the final step of the translation procedure to translate a `nice' FO3 formula into RA differs slightly.

For brevity, let $\dot{x}$ express $x \in P$, $\dot{y}$ express $y \in Q$, and $\dot{z}$ express $z \in R$, where $P$, $Q$, and $R$ are arbitrary elements in $\mathcal{D}$. In the heterogeneous setting, when we write $a(\dot{x}, \dot{y})$, for instance, this is a syntactic shorthand for the FO3 predicate $a(x, y)$, where $x \in P$ and $y \in Q$. This notation preserves the original FO3 syntax while making the type information explicit. The well-typedness condition introduced earlier remains: for every occurrence of a predicate $a(x, y)$, the types of $x$ and $y$ must match $d_1(a)$ and $d_2(a)$, respectively. Which means that for $a(\dot{x}, \dot{y})$, for instance, it must be true that $d_1(a)=P$ and $d_2(a)=Q$. Then, the heterogeneous translation function we implemented for the final step, $\text{\"T}$, is given as:
\begin{align*}
    &\begin{aligned}
        \text{\"T}([\true]_{\dot{x},\dot{y}}) &:= \full[P,Q] &
        \text{\"T}([\false]_{\dot{x},\dot{y}}) &:= \empty[P,Q] \\
        \text{\"T}([a(\dot{x}, \dot{y})]_{\dot{x},\dot{y}}) &:= a[P,Q] &
        \text{\"T}([a(\dot{y}, \dot{x})]_{\dot{x},\dot{y}}) &:= (a[P,Q])^{-1} \\
        \text{\"T}([a(\dot{x}, \dot{x})]_{\dot{x},\dot{y}}) &:= (a[P,P] \cap \identity[P,P]) \circ \full[P,Q] &
        \text{\"T}([a(\dot{y}, \dot{y})]_{\dot{x},\dot{y}}) &:= \full[P,Q] \circ (a[Q,Q] \cap \identity[Q,Q]) \\
        \text{\"T}([\neg(\varphi^{\{\dot{x}, \dot{y}\}})]_{\dot{x},\dot{y}}) &:= \overline{\text{\"T}([\varphi^{\{\dot{x}, \dot{y}\}}]_{\dot{x},\dot{y}})} &
        \text{\"T}([\dot{x} = \dot{x}]_{\dot{x},\dot{y}}) &:= \full[P,Q] \\
        \text{\"T}([\dot{x} = \dot{y}]_{\dot{x},\dot{y}}) &:= \identity[P,Q] &
        \text{\"T}([\dot{x} = \dot{y}]_{\dot{y},\dot{x}}) &:= (\identity[P,Q])^{-1} \end{aligned} 
        \end{align*}
        \begin{align*} \begin{aligned}
        \text{\"T}([\varphi^{\{\dot{x}, \dot{y}\}} \land \psi^{\{\dot{x}, \dot{y}\}}]_{\dot{x},\dot{y}}) &:= \text{\"T}([\varphi^{\{\dot{x}, \dot{y}\}}]_{\dot{x},\dot{y}}) \cap \text{\"T}([\psi^{\{\dot{x}, \dot{y}\}}]_{\dot{x},\dot{y}}) \\
        \text{\"T}([\varphi^{\{\dot{x}, \dot{y}\}} \lor \psi^{\{\dot{x}, \dot{y}\}}]_{\dot{x},\dot{y}}) &:= \text{\"T}([\varphi^{\{\dot{x}, \dot{y}\}}]_{\dot{x},\dot{y}}) \cup \text{\"T}([\psi^{\{\dot{x}, \dot{y}\}}]_{\dot{x},\dot{y}}) \\
        \text{\"T}([\exists \dot{z}. \varphi^{\{\dot{x},\dot{z}\}} \land \psi^{\{\dot{z},\dot{y}\}}]_{\dot{x},\dot{y}}) &:= \text{\"T}([\varphi^{\{\dot{x}, \dot{z}\}}]_{\dot{x},\dot{z}}) \circ \text{\"T}([\psi^{\{\dot{z},\dot{y}\}}]_{\dot{z},\dot{y}}) \\
        \text{\"T}([\forall \dot{z} . \varphi^{\{\dot{x},\dot{z}\}} \lor \psi^{\{\dot{z},\dot{y}\}}]_{\dot{x},\dot{y}}) &:= \text{\"T}([\varphi^{\{\dot{x}, \dot{z}\}}]_{\dot{x},\dot{z}}) \dagger \text{\"T}([\psi^{\{\dot{z},\dot{y}\}}]_{\dot{z},\dot{y}})
    \end{aligned}
\end{align*}
Just as was done for the homogeneous translation, the notation being used here for the heterogeneous translation function is $\text{\"T}([\varphi]_{a,b})$ where $\varphi$ is the FO3 formula to translate, and $a$ and $b$ are two arbitrary variables that are used to keep track of the source and target of the relation algebra formula being built by the translation. In our code~\cite{doi_of_implementation}, the translation function is thus implemented to accept three arguments: $\varphi$, $a$, and $b$. Again, setting $a$ and $b$ to be pairwise distinct ensures that the definition does not become ill-defined in the case of atomic formulas and equality. Since $a$ and $b$ can be arbitrary in the initial call to the translation function, we typically set them as $x \in Left$ and $y \in Right$, respectively.

In the last two lines of the translation function definition, $\varphi^{\{\dot{x}, \dot{z}\}}$ and $\psi^{\{\dot{z}, \dot{y}\}}$ stand for the conjuncts (disjuncts) that depend on the variables $\{\dot{x},\dot{z}\}$ and $\{\dot{z},\dot{y}\}$ inside the existential (universal) quantifier, respectively.

To demonstrate the heterogeneous translation procedure, a fully worked example of the translation of the closed FO3 formula $\forall \dot{x}. \forall \dot{y}. \exists \dot{z}. (\neg((A(\dot{x},\dot{z})) \land (B(\dot{z},\dot{x})))) \land (C(\dot{x},\dot{y}))$ into RA is shown below. The translation function $\text{\"T}([\forall \dot{x}. \forall \dot{y}. \exists \dot{z}. (\neg((A(\dot{x},\dot{z})) \land (B(\dot{z},\dot{x})))) \land (C(\dot{x},\dot{y}))]_{x \in Left, y \in Right})$ is used to produce the final translation, which is why the types ``Left" and ``Right" appear in the final translation.

\begin{enumerate}
    \item Original Expression: $\forall \dot{x}. \forall \dot{y}. \exists \dot{z}. (\neg((A(\dot{x},\dot{z})) \land (B(\dot{z},\dot{x})))) \land (C(\dot{x},\dot{y}))$
    \item Negation Normal Form: $\forall \dot{x}. \forall \dot{y}. \exists \dot{z}. ((\neg(A(\dot{x},\dot{z}))) \lor (\neg(B(\dot{z},\dot{x})))) \land (C(\dot{x},\dot{y}))$
    \item Good FO3 Translation: $\forall \dot{x}. \forall \dot{y}. (\exists \dot{z}. (\neg(A(\dot{x},\dot{z}))) \land (C(\dot{x},\dot{y}))) \lor (\exists \dot{z}. (\neg(B(\dot{z},\dot{x}))) \land (C(\dot{x},\dot{y})))$
    \item Nice FO3 Translation: $\forall \dot{x}. \forall \dot{y}. ((C(\dot{x},\dot{y})) \land (\exists \dot{z}. \neg(A(\dot{x},\dot{z})))) \lor ((C(\dot{x},\dot{y})) \land (\exists \dot{z}. \neg(B(\dot{z},\dot{x}))))$
    \item Final Translation: $(\empty[Left,P]) \dagger ((((((C[P,Q]) \cap ((\overline{A[P,R]}) \circ (((T[R,R]) \cap (\identity[R,R])) \circ (T[R,Q])))) \cup ((C[P,Q]) \cap ((\overline{(B[R,P])^{-1}}) \circ (((T[R,R]) \cap (\identity[R,R])) \circ (T[R,Q]))))) \dagger (((\empty[Q,Q]) \cup (\overline{\identity[Q,Q]})) \dagger (\empty[Q,P]))) \cup (\overline{\identity[P,P]})) \dagger (\empty[P,Right]))$
    \item Final Translation Simplified: $(\empty[Left,P]) \dagger ((((((\overline{((B[R,P])^{-1}) \cap (A[P,R])}) \circ (\full[R,Q])) \cap (C[P,Q])) \dagger (\empty[Q,P])) \cup (\overline{\identity[P,P]})) \dagger (\empty[P,Right]))$
\end{enumerate}

\section{Implementation process}
Our implementation is written in Python 3.11 and can be found online~\cite{doi_of_implementation}. Please refer to the included README.md file for setup and usage instructions. We used three important principles to guide our development: make all code testable early, write single-assignment code, and generate code that cannot reasonably be fully covered with tests.

In creating our tool, our first focus was making our code testable. For this, we implemented a translation of homogeneous relation algebra to first-order logic, following the interpretation function. We also implemented a - straightforward - translation of first-order logic (in our data structures) into the structures used to represent them in Z3. From this point on all code we wrote could be tested, as outlined in the next section.

As we moved toward heterogeneous translation, we added run-time checks on the datatype for all constructions. Whenever an RA formula is constructed, we test at the top level whether it is well-typed. Section~\ref{sec:typechecking} says a few words about this.

Another principle we used was to write as much single-assignment code as possible. The transformation of FO3 formulas into `good' and `nice' FO3 formulas does not (or should not) change the semantics, so one might be tempted to go with a memory-efficient implementation that replaces objects representing formulas by their equivalent `good' and `nice' versions. We believe this benefit to be quite minor, as code that manipulates formulas is hardly ever part of a critical loop that needs to be heavily optimized. For our use case, the code would be part of a parser of user-written scripts, where the benefit would be negligible. A major downside of re-assignments is they make code much harder to reason about. Unintentional re-assignments did occur at some point and led to bugs (that were immediately caught by our testing). This was due to lists being mutable in Python, a cause that was difficult to spot.

For the most part, our implementation follows the procedure outlined in Section~\ref{translation}, much of which we implemented as recursive pattern matching on the formula given. All four steps mentioned in the previous section are contained within our translation tool. The resulting formulas are sometimes overly verbose. Rather than fixing this by adding special cases to our translation, we added a simplification step to clean up the RA formulas produced. Our simplifier looks at all sub-formulas of a formula in RA, and checks if they can be simplified. If so, the simplifier is called again on the simplified formula. A problem with writing code for this is that many formulas can be simplified, leading to verbose code and a big potential for errors. Moreover, random testing might miss certain formulas, especially if a bug is caused by the interplay of several steps. As a consequence, we decided not to write the simplifier by hand, but to generate the code for it instead. This is illustrated in Section~\ref{sec:simplification}.

\subsection{Validation with Z3}
To ensure the correctness of our implementation, we conducted automated testing using the z3-solver Python module. Our validation process involves the following steps:

\begin{enumerate}
\item \textbf{Random Formulas Generation:} We utilize a method we created to generate random closed FO3 formulas. This method was written specifically for validation purposes. It can be given parameters to specify the size of the formula to generate as well as how many of them to generate.
\item \textbf{Translation to RA:} The generated formulas are translated into RA using our implementation of Nakamura's procedure.
\item \textbf{Simplification of RA:} We run tests with and without the simplifier. While building the translations initially, there was no simplifier. Since the simplifier should not change the semantics of RA formulas, this step is optional for the validation.
\item \textbf{Translation Back:} The RA formulas are translated back into FO3. This is the standard direction of the translation, and while it is not necessary to implement for our tool, it's useful to have it specifically for validation.
\item \textbf{Equivalence Check:} Finally, we use Z3 to verify if the original formula and the output of the previous step are equivalent. To do so, we ask Z3 to find a satisfying assignment to the statement that the formulas are not equivalent. If Z3 finds such a counterexample, it indicates an error in our implementation.
\end{enumerate}

To create a flexible and robust testing system, we generated random closed FO3 formulas of various specified sizes (number of symbols). This approach enabled us to validate our code thoroughly. In practice, our validation method helped us discover and eliminate initial bugs present in early versions of our software tool.

Adapting our software tool to include support for translating formulas in the heterogeneous setting became significantly easier with the assistance of Z3. Leveraging Z3, we were able to employ a trial-and-error approach for some parts of the work by initially making educated guesses about the code, testing it with Z3, and subsequently analyzing the results to determine the correctness or incorrectness of the written code. This facilitated a more effective development process.

We observed that for many equivalent formulas, Z3 can prove their equivalence very quickly. To use this, we run Z3 with a low timeout initially (about 60 ms). If Z3 times out and returns `unknown', we have a high chance that the formula is false, and we rerun the query on a finite sort with four elements and a larger timeout (about 1s): Many non-equivalent formulas are quickly disproved with a small finite model. In this case, if Z3 returns `unsat' (it was unable to find a counterexample) or `unknown', we rerun the query, using a very large timeout (minutes). This last case rarely occurs in practice.

\subsection{Type Checking}\label{sec:typechecking}
Upon creation of heterogeneous RA objects, we check to ensure that the properties of a well-typed formula defined in Section~\ref{background} hold. If the properties hold, then the RA formula is well-typed. If one (or more) of the properties is broken, however, then our code raises an exception to alert the user of our tool that an ill-typed RA formula has been created. Errors caught using this check were minor typos that were easy to fix. It is hard to say whether they would have been as easy to fix if these checks were not in place.

While an end-user should never have to experience such exceptions while using our tool, the constructors of RA formulas are considered `public', so we left these safety checks in place. This way, users who use these constructors directly will be alerted of ill-typed formulas if they occur.

\subsection{Simplification of RA Formulas}\label{sec:simplification}
Building upon the concepts used in our automated testing implementation, we also developed a reliable tool for simplifying RA formulas. The code to simplify RA formulas is generated by a separate Python program.

The first step in building our simplifier was to enumerate all formulas that can be considered simplifications for homogeneous RA formulas. To do so, we wrote code that could generate all formulas of a specific size, similar to the code that randomly creates formulas of a certain size. Here we used the \verb|yield| keyword in Python to generate all possible formulas of that size. This produces an iterator that generates all formulas, and the resulting code looks similar to what one would write using a list monad for this in a functional programming language.

To come up with all simplifying rules, we generate pairs of RA formulas $(\varphi, \psi)$, where $\varphi$ is of a specified size and $\psi$ is smaller than it. Moreover, we require that relation symbols in $\psi$ occur at least as often in $\varphi$. As a consequence, replacing a formula $\varphi$ by $\psi$ in the way of formula rewriting would reduce the size of the overall formula. From the generated rules, we select only those that are valid by checking their equality with Z3. The valid rules are then stored in a dictionary which is saved to a file for future use.

We generate all simplifying rules starting at size 1 and increase this size until generating all rules takes too much time. To reduce the number of rules that need to be tested, we require that $\varphi$ and $\psi$ cannot be simplified by any rules generated in earlier iterations. We call the simplifier on $\varphi$ and $\psi$ to check this.

We generate Python code to simplify RA formulas from the dictionary of simplification rules. The generation process begins by writing code that matches $\varphi$ and replaces it with $\psi$. An example is given in Figure~\ref{fig:simplification}. The code is generated such that blocks with similar prefixes can be grouped, and by doing so the resulting code is more efficient (because patterns are only matched once) and concise. The grouping is done by finding common string prefixes using a function (\textbf{`group\_by\_prefix'}), after which it is written to a file (by \textbf{`write\_grouped\_code'}).

% Create Figure 2
\begin{figure}[htbp]
    \centering
    \begin{lstlisting}
def simplify(formula):
    if isinstance(formula, COR_Expressions.Intersection):
        lhs1, rhs1 = formula.argument1, formula.argument2
        if isinstance(lhs1, COR_Expressions.Union):
            lhs2, rhs2 = lhs1.argument1, lhs1.argument2
            A = lhs2
            B = rhs2
            if str(B) == str(rhs1):
                return ($``((A) \cup (B)) \cap (B) = B"$, B)
    \end{lstlisting}
    \caption{Simplification code generated for the rule $((A) \cup (B)) \cap (B) = B$}
    \label{fig:simplification}
\end{figure}

The generated simplifier matches the formula at the top level and returns the simplified formula along with the rule used to perform the simplification. This is done so we can use the simplifier to eliminate redundant rules and dead code in the generated files. For instance, the rules $A \cap A \rightarrow A$ and $B \cap B \rightarrow B$ both match on precisely the same formulas. However, when $B \cap B$ is provided to the simplifier, the returned rule is $A \cap A \rightarrow A$, indicating that the rule $B \cap B \rightarrow B$ is redundant.

To find all RA rules, we observe that any valid well-typed rule remains valid if all domains are set to $\mathcal{U}$. This means that every RA rule is represented by one in the homogeneous setting. To find all heterogeneous rules, we substitute the $\mathcal{U}$ for arbitrary types and test if the result is well-typed. If the resulting rule is well-typed, another equivalence check is performed in Z3 to ensure the rule is still valid. The resulting rules are then run through the simplifier to remove redundant rules.

Table~\ref{tab:homogeneous_simplification} and Table~\ref{tab:heterogeneous_simplification} below provide small samples of the homogeneous and heterogeneous RA simplification rules we have identified, respectively. For a complete list of all simplification rules identified by our method, please consult the files \texttt{COR\_Rules.txt} and \texttt{Typed\_COR\_Rules.txt} included in our 
implementation~\cite{doi_of_implementation}.

\begin{table}[htbp]
    \centering
    \begin{tabular}{|c|c|}
        \hline
        \textbf{Original Formula} & \textbf{Simplified Formula} \\
        \hline
        $A \cup A$ & $A$ \\
        $\identity^{-1}$ & $\identity$ \\
        $(A \cup B) \cup B$ & $B \cup A$ \\
        $\overline{A} \cup A$ & $\full$ \\
        $(A^{-1})^{-1}$ & $A$ \\
        $A \cap \overline{A}$ & $\empty$ \\
        \hline
    \end{tabular}
    \caption{Sample of Homogeneous RA Simplification Rules}
    \label{tab:homogeneous_simplification}
\end{table}

\begin{table}[htbp]
    \centering
    \begin{tabular}{|c|c|}
        \hline
        \textbf{Original Formula} & \textbf{Simplified Formula} \\
        \hline
        $A[P,Q] \cup A[P,Q]$ & $A[P,Q]$ \\
        $\identity[P,P]^{-1}$ & $\identity[P,P]$ \\
        $(A[P,Q] \cup B[P,Q]) \cup B[P,Q]$ & $B[P,Q] \cup A[P,Q]$ \\
        $\overline{A}[P,Q] \cup A[P,Q]$ & $\full[P,Q]$ \\
        $(A[P,Q]^{-1})^{-1}$ & $A[P,Q]$ \\
        $A[P,Q] \cap \overline{A}[P,Q]$ & $\empty[P,Q]$ \\
        \hline
    \end{tabular}
    \caption{Sample of Heterogeneous RA Simplification Rules}
    \label{tab:heterogeneous_simplification}
\end{table}

\subsection{Note on our Use of Z3}
The role of Z3 in this work is threefold:

\begin{enumerate}
    \item We utilize Z3 for code validation during testing.
    \item Z3 aids in code generation by identifying and removing incorrect code.
    \item Z3 serves as a benchmark to illustrate the utility of relation algebra as an alternative to first-order logic reasoning.
\end{enumerate}

Importantly, none of these roles necessitate end users of our tool to run Z3 directly. Our developed tool incorporates results derived from Z3 rather than requiring its direct invocation.

Z3 is used to validate our translation tool, and conversely, we can use our translation tool to evaluate Z3's effectiveness as a theorem prover. During testing, we encountered equations that Z3 struggled to prove or disprove within a reasonable time. Z3 can prove these equations for a finite sort, but it otherwise times out after 1 minute. Typically, these equations are quite large and complex. The following is an example of one such equation:

\[\forall y. \forall z. (\exists x. (\exists y. (B(y, x)) \land (B(x, z))) \lor (\neg(\neg(C(x, y))))) \land ((C(y, y)) \lor (\true))\]
\[\stackrel{?}{=}\]
\[(\forall y. \exists z. (B(z, y)) \land (\exists y. B(y, z))) \lor (\forall z. \exists y. C(y, z))\]

It is significant that Z3 (version 4.12.1 of the z3-solver Python library) times out on this question because the question is an FO3 formula. Translating the equation to RA using our simplifier results in a trivially true equation, accomplished within milliseconds. This highlights potential areas for improvement within Z3 itself by leveraging our RA translation. Moreover, it indicates that combining Z3 and tools like it with a tool that performs equational reasoning in RA has the potential to significantly enhance Z3's capabilities.

To further illustrate the merit of RA simplification, consider another example of an FO3 equation where Z3 times out after 1 minute when attempting to prove or disprove the given FO3 equation, whereas translating this equation to RA and simplifying it using our simplifier results in a trivially true equation:

\[\forall x. \forall y. (\true) \land (\exists z. (((A(y,y)) \lor (\false)) \lor (\neg(\neg(\neg((\exists y. z = x) \land (\true)))))) \land (\neg(((y = y) \land (\false)) \lor (A(y,z)))))\]
\[\stackrel{?}{=}\]
\[\forall z. \forall y. (\neg(\forall x. (z = x) \lor (A(y,x)))) \lor ((\exists z. (A(y,z)) \land (y = z)) \land (\exists z. \neg(A(y,z))))\]

\subsection{Deviations from the Original Procedure}\label{deviations}

The translation, as originally given~\cite{Nakamura2020} and as presented in this paper, discusses FO3 as a language where only three variables can occur. When we introduced FO3, $x$ and $y$ were variables taken from a set with precisely three elements. In practice, the creator of an FO3 formula might prefer to use meaningful variable names. 
To allow this, we have a slightly lighter restriction which is that we require at most three unbound variables to occur under every quantifier, allowing us to rename variables to get an FO3 formula.
Therefore, our implementation translates first-order logic formulas for which there is a straightforward correspondence to an FO3 formula. Consider the following first-order logic formula as an example:
\[\exists x \in A. \exists y \in B. \exists z \in C. \exists w \in D. ~a(x, w)\]

Its translation would be:
\[\full_{[Left,A]}\circ a \circ \full_{[D,right]}\]

The translation function $\text{\"T}([\exists x \in A. \exists y \in B. \exists z \in C. \exists w \in D. ~a(x, w)]_{x \in Left, y \in Right})$ is used to produce this translation, which is why the types ``Left" and ``Right" appear in the translation.

\subsection{Scope and Metrics}
The code accompanying this work encompasses 5410 lines, composed of 2338 lines typed by hand and 3072 lines of simplification code generated through automated processes using Z3. Of these 5410 combined lines of code, 4254 are part of the final product that an end user would interact with, while the remaining 1156 lines were used only to create the final product.

Another interesting statistic is that our rule dictionaries currently contain 250 homogeneous simplification rules and 263 heterogeneous simplification rules, meaning 13 of our heterogeneous rules could not be generalized by another typing of the same rule. For example, the following are two different typings found in our heterogeneous rules dictionary of the same simplification rule, with neither one generalizing the other:
\[((A[P, P]) \circ (B[P, Q])) \cup ((A[P, P]) \circ (C[P, Q])) = (A[P, P]) \circ ((B[P, Q]) \cup (C[P, Q]))\] 
\[\text{and}\] 
\[((A[P, Q]) \circ (B[Q, Q])) \cup ((A[P, Q]) \circ (C[Q, Q])) = (A[P, Q]) \circ ((B[Q, Q]) \cup (C[Q, Q]))\]

\section{Conclusion and Future Work}

This work implemented a translation from FO3 to RA, and back. We addressed a translation to homogeneous relation algebra, as well as typed FO3 formulas and a translation to heterogeneous Relation Algebra, which to the best of our knowledge is a new contribution. Furthermore, we extensively tested our implementation using Z3 and our type-checking system.

Some details still need to be figured out before people can write First-Order Logic formulas as an alternative to writing valid Ampersand or RelView code. These details include minor syntax issues and the fact that neither Ampersand nor RelView is built in Python. In the case of Ampersand~\cite{Joosten2015}, the heterogeneous Relation Algebra used allows for sub-typing: if A is a subset of B, then a relation R from A to A can be composed with a relation S from B to B. While this should relax the language in principle, we need to confirm that translations in these cases are as expected.

A second question is how to get around the issue of only three variables being allowed in FO3 formulas. While three variables suffice for many declarations, one can occasionally hit this limit, which might surprise the unwary user. Introducing additional predicates can help get around this limit, as shown by Givant~\cite{Givant2006}. Doing this practically means that we do not wish to introduce too many predicates. Moreover, we wish their values could be calculated reasonably efficiently in systems like Ampersand or RelView. How to achieve this is an open question.

Furthermore, we have discovered through our work that there are some FO3 equations that Z3 is unable to prove nor disprove the validity of in a reasonable amount of time. It would be interesting to look further into why this is and how Z3 could be improved to prove or disprove such equations quickly.

\bibliography{citations}

\begin{thebibliography}{1}
\providecommand{\url}[1]{\texttt{#1}}
\providecommand{\urlprefix}{URL }
\expandafter\ifx\csname urlstyle\endcsname\relax
  \providecommand{\doi}[1]{doi:\discretionary{}{}{}#1}\else
  \providecommand{\doi}{doi:\discretionary{}{}{}\begingroup
  \urlstyle{rm}\Url}\fi
\providecommand{\eprint}[2][]{\url{#2}}

\bibitem{Berghammer2005}
Berghammer R, Neumann F.
\newblock RelView--An OBDD-based Computer Algebra system for relations.
\newblock In: International Workshop on Computer Algebra in Scientific
  Computing. Springer, 2005 pp. 40--51.

\bibitem{doi_of_implementation}
Brogni A, Joosten SJC.
\newblock Translating First-Order Predicate Logic to Relation Algebra.
\newblock \textit{Available online at
  \url{https://doi.org/10.5281/zenodo.14285122}}, 2024.

\bibitem{Givant2006}
Givant S.
\newblock The calculus of relations as a foundation for mathematics.
\newblock \emph{Journal of Automated Reasoning}, 2006.
\newblock \textbf{37}(4):277--322.

\bibitem{Joosten2015}
Joosten SMM, Joosten SJC.
\newblock Type checking by domain analysis in ampersand.
\newblock In: International Conference on Relational and Algebraic Methods in
  Computer Science. Springer, 2015 pp. 225--240.

\bibitem{z3_theorem_prover}
{Microsoft Research}.
\newblock Z3, An efficient SMT solver.
\newblock \textit{Available online at
  \url{https://github.com/Z3Prover/z3/tree/z3-4.12.1}}, 2023.

\bibitem{Nakamura2020}
Nakamura Y.
\newblock Expressive Power and Succinctness of the Positive Calculus of
  Relations.
\newblock In: International Conference on Relational and Algebraic Methods in
  Computer Science. Springer, 2020 pp. 204--220.

\end{thebibliography}
\end{document}